\documentstyle[11pt,newpasp,twoside,epsf,graphicx]{article}

\begin{document}

\title{Formation and pre--MS evolution of massive stars with growing accretion}
\author{A. Maeder, R. Behrend}
\affil{Geneva Observatory, CH-1290 Sauverny, Switzerland}

\begin{abstract}
We briefly describe the three existing scenarios for forming massive stars and emphasize
that the arguments often used to reject the accretion scenario for massive 
stars are misleading. It is usually  not accounted for the fact
that the turbulent pressure associated to large turbulent velocities in clouds necessarily imply 
relatively high accretion rates for massive stars.

We show the basic difference between the formation of low and high mass stars based on the values of
the free fall time and of the Kelvin-Helmoltz timescale, and define the concept of birthline
for massive stars.

Due to  D-burning, the radius and  location of the birthline in the HR diagram, as well as
the lifetimes are very sensitive to the accretion rate $\dot M_{\mathrm{accr}}$. If a form
$\dot M_{\mathrm{accr}} \propto A\,(M/M_{\sun})^\varphi$ is adopted, the observations in
the HR diagram and the lifetimes support a value of $A \approx
10^{-5}\,M_{\sun} \cdot {\mathrm{yr}}^{-1}$ and a value of $\varphi \ga 1$. Remarkably, such a
law is consistent with the relation found by Churchwell (1998) and Henning et al. (2000)
between the outflow rates and the luminosities of ultra--compact HII regions,
if we assume that a fraction $0.15$ to $0.3$ of the global inflow is accreted.
The above relation implies high $\dot M_{\mathrm{accr}} \approx 10^{-3}\, M_{\sun} \cdot {\mathrm{yr}}^{-1}$
for the most massive stars. The physical possibility of such high $\dot M_{\mathrm{accr}}$ is
supported by current numerical models.

Finally, we give simple analytical arguments in favour of the growth of $\dot M_{\mathrm{accr}}$ with the
already accreted mass. We also suggest that due to Bondi-Hoyle accretion, the formation of
binary stars is largely favoured among massive stars in the accretion scenario.
\end{abstract}

\section{Introduction}

Numerous new observations in radio, IR, optical, 
UV and X--rays, together with
many numerical models, have  contributed
to the progress of the field of star formation. However,
the formation of massive stars is still a major unsolved problem in stellar
evolution. 

The answer brought to the question "How do massive stars form ?" has a major
impact not only for stellar evolution, but also for spectral and chemical evolution of galaxies
as well as for cosmology.
Several competing processes have now been identified leading to different
scenarios for the formation of massive stars. Nevertheless, this does not mean that the
authors of these scenarios are competing. Rather they all look for the 
proper answer to the above major question.

\section{The various scenarios for the formation of massive stars}

One can identify three different scenarios:\\

\indent a) The classical scenario.\\
This is the pre--MS evolution at constant mass with bluewards horizontal tracks in the HR
diagram, moving from the Hayashi line to the zero age 
main sequence (ZAMS), ({\it cf}. Iben 1963).
The timescale is the Kelvin-Helmholtz timescale $t_{\mathrm{KH}} \approx G\, M^2/R\, L$ which is
about 0.5 to 2\% of the MS lifetime. For example, $t_{\mathrm{KH}}$ 
is about $3 \cdot 10^4\,{\mathrm{yr}}$ for a $30\,M_{\sun}$ star.
Due to lots of evidence of mass accretion (see below, \S~\ref{evidence}), this scenario is
no longer supported, although it is a good reference basis.\\

\indent b) The collision or coalescence scenario.\\
Protostars are moving around in a young cluster and collisions of intermediate mass
protostars may lead to the formation of massive stars. This interesting possibility and
its consequences have been extensively studied by Bonnell, Zinnecker and colleagues ({\it cf}.
Bonnell et al. 1998).

Often in the literature ({\it cf}. Stahler 1998), it is said that the coalescence scenario is
necessary, on the basis of the argument
 that the accretion scenario is not possible for massive stars because
their high radiation pressure on the dust may reverse the infall and prevent the accretion.
We emphasize that the coalescence scenario may well be important, but not for this specific reason.
The reasons are the following ones.

We may consider that the accretion rate $\dot M_{\mathrm{accr}}$ is given by
\begin{equation}
   \dot M_{\mathrm{accr}} \approx c_{\mathrm{s}}{}^3/G 
   \label{form1}
\end{equation}
as resulting from the ratio of the Jeans mass by the free fall time. For
typical temperatures of $10$-$100\,{\mathrm{K}}$, as observed in molecular clouds,
a value of 
$\dot M_{\mathrm{accr}} \approx 10^{-5}\,M_{\sun} \cdot {\mathrm{yr}}^{-1}$ is  obtained from
relation~(\ref{form1}). It is
true that for such values of $\dot M_{\mathrm{accr}}$, the effects of the radiation pressure
$P_{\mathrm{rad}}$ can reverse the infall. However, this
argument ignores the role of turbulent pressure in the clouds. Indeed, 
high turbulent velocities have been
observed in regions of massive stars ({\it cf}. Tatematsu et al. 1993; Caselli \&
Myers 1995; Nakano et al. 1995; Nakano et al. 2000). If we account for the
turbulent velocities in equation~(\ref{form1}), there is more support in the clouds
which are thus denser and one gets much higher accretion rates
of the order of 
\begin{displaymath}
   \dot M_{\mathrm{accr}} \approx 10^{-3}\,M_{\sun} \cdot {\mathrm{yr}}^{-1}.
\end{displaymath}

\noindent
For such a high accretion rate, the momentum of the infalling material is higher than the
momentum in the outgoing radiation of a massive star and the accretion can not be reversed
(see below, \S~\ref{faisable}). Thus, the basic argument often used
against the accretion scenario is invalid, since it does not account for
the fact that the accretion rates are very high.

This does not mean that the coalescence scenario never works. But, one has
to be careful about the question of its timescale. In particular Elmegreen
(2000) has suggested that "there is not enough time for a protostar to
move around in a young cluster and to coalesce with other protostars".
Certainly collisions do occur, but the real importance of this scenario
has still to be ascertain.\\

\indent c) The accretion scenario.\\
The accretion scenario with constant $\dot M_{\mathrm{accr}}$ was first proposed by Beech \&
Mitalas (1994). It is however evident that with moderate accretion rate
like $\dot M_{\mathrm{accr}} \approx 10^{-5}\,M_{\sun} \cdot {\mathrm{yr}}^{-1}$, we would need
$10^7\,{\mathrm{yr}}$ to form a $100\,M_{\sun}$ star. At this age, it would already have
exhausted its central hydrogen. Thus, one needs accretion rate growing with time or with
the stellar mass ({\it cf}. Bernasconi \& Maeder 1996; Norberg \& Maeder 2000;
Behrend \& Maeder 2001;
McKee, this meeting). The various properties of these models will be
discussed in section~\ref{modeles}.

\section{Brief summary of the observational evidence for the formation of massive stars
by accretion \label{evidence}}

As shown by several recent works as well as in this meeting,
there are several observational indications in favour of the massive star formation by
accretion.

\begin{itemize}
\item{Massive outflows.}

There is evidence from radio observations of massive outflows ({\it cf}. Churchwell
1998; Henning et al. 2000) with outflow rates $\dot
M_{\mathrm{accr}} \propto L^{0.75}$ where $L$ is an estimate of the stellar luminosity. 
A fraction $15\%$ to
$30\%$ of the global infall is supposed to be accreted, while the rest goes
into the outflows. This
suggests that during the formation of massive stars, the accretion rate may grow
relatively steeply with the already accreted stellar mass.

\item {Velocity dispersion.}

High velocity dispersions are observed, as for example from the CS line in the Orion KL
Nebula (Caselli \& Myers 1995).
This supports the idea that turbulence provides enough support to the cloud 
to allow a high enough density, which then leads to accretion rates of the order of
$10^{-3}\,M_{\sun} \cdot {\mathrm{yr}}^{-1}$.

\item{Luminosity.}

The luminosity of IR sources like the Orion IRC2  K--luminosity is quite compatible with
luminosities of $L \propto {G\,M\,\dot M \over R}$ of discs having accretion rates of the order
of $10^{-2}$ to $10^{-3}\,M_{\sun} \cdot {\mathrm{yr}}^{-1}$ (Morino et al. 1998;
Nakano et al. 2000).

\item{Spectra.}

Also the spectral distribution of the energy emitted by hot cores in hot molecular clouds
is compatible with discs having accretion rates as given above (Osorio et al. 1999).

\item{Direct evidence for discs.}

Direct evidence for discs has been provided. An example is IRAS 20126 +4004 from
$\textrm{NH}_3$
line observed by the VLA (Zhang et al. 1998; Cesaroni 2000).
The disc looks perpendicular to the molecular outflow, and a velocity gradient is observed
in the disc. Further compelling evidence of discs has been presented at this meeting by Zhang
and by Conti.

\end{itemize}

\noindent
Interestingly enough, a prediction of the coalescence scenario is that discs should 
be destroyed by the
collision, thus the visibility of discs may be an evidence against the coalescence.

\section{The accretion scenario: semi-empirical approach \label{modeles}}

\subsection{Generalities}

We are using various approaches, here. The semi-empirical approach which gives rough overall
constraints, the analytical one which emphasizes the relation between  basic parameters, and the
accurate numerical one ({\it cf}. \S~\ref{depend}).

\begin{figure}[!htb]
   \plotone{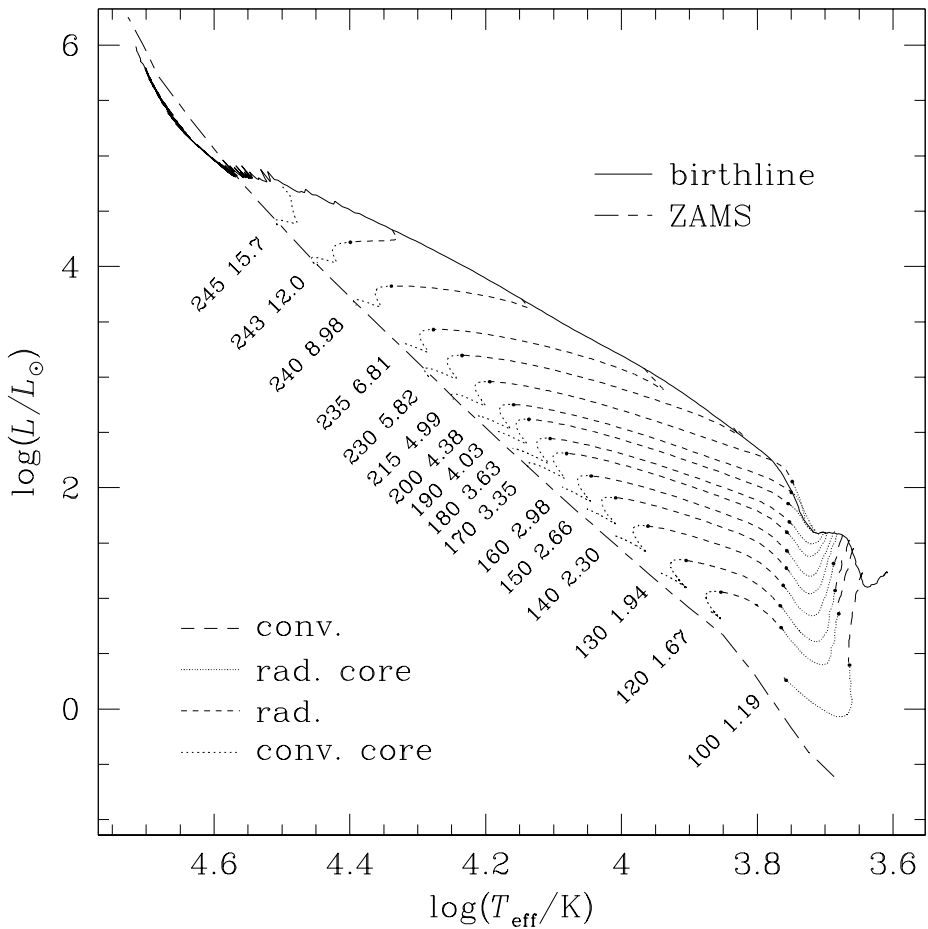}
   \caption{Evolutionary tracks with accretion rate given by equation
(6) with
   $\tilde f=0.5$.
   The long and short dashed line is the ZAMS.
   Firstly the stars are fully convective, then radiative cores are developed and grow
   in size until a temporary fully radiative state is attained,
   the stars have finally a convective core.
   The age and mass of the stars when they quit the birthline are indicated
   in units of $10^3\,{\mathrm{yr}}$ and $M_{\sun}$ at the 
end of the individual tracks, from Behrend \& Maeder (2001)}
   \label{diag10}
\end{figure}

There is a fundamental difference between low and high mass stars, from the point of view
of star formation if we consider the free fall time
\begin{equation}
   t_{\mathrm{ff}}={3\,\pi \over \sqrt{32\,G\,\bar \rho_{\mathrm{in}}}}
   \label{form2}
\end{equation}
and the Kelvin-Helmoltz timescale
\begin{equation}
   t_{\mathrm{KH}}={G\,M^2 \over R\,L} \, ,
   \label{form3}
\end{equation}
where $\bar \rho_{\mathrm{in}}$ is the average initial density of the cloud supposed to be
at Jeans limit ({\it i.e.}~$R_{\mathrm{J}} \propto M_{\mathrm{J}})$.
For $M \leq 8\,M_{\sun}$ (in the numerical model of Figure~\ref{diag10}, this limit turns
to be around $15\,M_{\odot}$), one has 
\begin{displaymath}
   t_{\mathrm{ff}}<t_{\mathrm{KH}}.
\end{displaymath}
This means that the accretion process is completed long before the central contraction has
initiated  nuclear burning. Thus, in this range of masses, 
the evolution is not so far
from the evolution at constant mass. For $M \ga 8\,M_{\sun}$, one has  
\begin{displaymath}
   t_{\mathrm{ff}}>t_{\mathrm{KH}}.
\end{displaymath}
In this case, the accretion is not yet completed when the central core has already
finished its contraction and started nuclear burning. Thus, 
 massive stars start their
H-burning largely hidden in their molecular clouds, a fact consistent with the
observations by Wood \& Churchwell (1989).

Figure~\ref{diag10} illustrates several properties of the evolution with accretion. We start
the evolution from a $0.7\,M_{\sun}$ star which is accreting
 at the indicated rate
$\dot M_{\mathrm{accr}}$.
The birthline is the path in the HR diagram followed by a star accreting at a
specified rate, significant for dominating the evolution. If at some stage on the birthline the
accretion is stopped, then the star sets on an individual, 
more horizontal, track
corresponding to its mass and moving to the ZAMS. 
Certainly, in reality, accretion does not stop abruptly,
but progressively with some exponential decline. However, if the decrease is fast enough, this
will make no difference. The star finally reaches the ZAMS 
following a track not too different from the 
track at constant mass. However, the timescale on the tracks can be different from
those with constant mass evolution.
The accretion rate used in Figure~\ref{diag10}
is basically one half of the outflow rate given by Churchwell (1998) and Henning et al. (2000).
      
We notice in Figure~\ref{diag10} that the time required to form massive stars of different masses are
not very different, as for example for the models of $12$ and $15.7\,M_{\sun}$. This
directly results from the accretion law used to construct the birthline.
The birthline joins the ZAMS near $8\,M_{\sun}$ for low accretion rate of the order of
$10^{-5}\,M_{\sun} \cdot {\mathrm{yr}}^{-1}$ 
({Palla \& Stahler 1993), while for the
higher accretion rate used in Figure~\ref{diag10}, it joins the ZAMS near $15\,M_{\sun}$.
Above this critical point (which depends on $\dot M_{\mathrm{accr}}$), the birthline more
or less coincides with the ZAMS. This means that the massive accreting stars  are ascending the
ZAMS, as they continue their accretion. This is a new kind of a pre--MS track: the upward
track along the ZAMS of a rapidly accreting star. Of course, if the accretion rate does
not keep high enough, we  may have a progressive evolutionary displacement along post-MS tracks. 

There are several differences between the present  models on the ZAMS, due
to their particular history, and models which would be just started homogeneously
on the ZAMS ({\it cf}. Bernasconi \& Maeder 1996).
1) A newly formed massive star with $M>40\, M_{\sun}$ at the time it emerges from its cloud
has already burnt several percents of its hydrogen.
2) A proper ZAMS does thus not exists for massive stars.
3) The part of the MS lifetime, during which the star is visible,
is thus  reduced.
4) The size of the convective cores are 10\% smaller than for homogeneous classical ZAMS stars.

\begin{figure}[!htb]
   \plotone{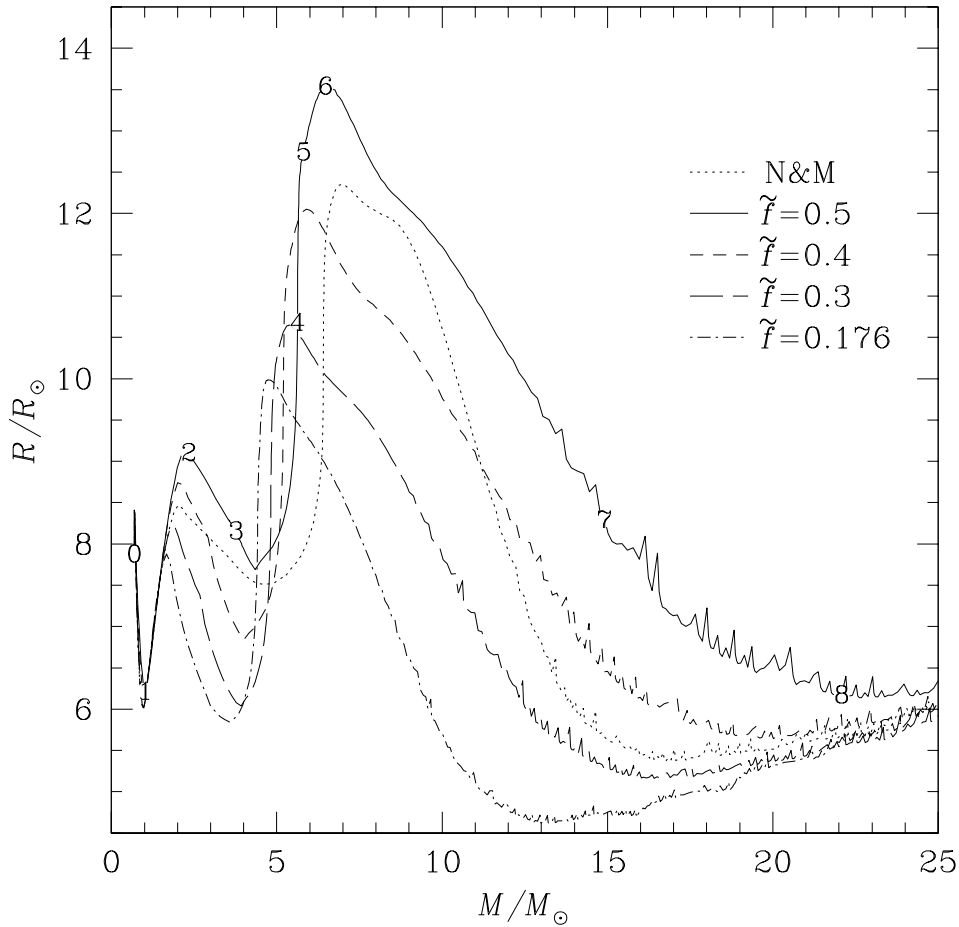}
   \caption{Mass vs. radius relation for stars on the birthline. The curves labeled
      $\tilde f=0.5,\,0.4,\,0.3,\,0.176$ correspond to accretion rate which are equal to
      $\tilde f$ time the
      outflow rates given by Churchwell (1998) and Henning et al.
      (2000), see equation (6). The track N\&M is from Norberg \& Maeder (2000).
      Points 1: central ignition of deuterium, 3: start of shell
      D-burning, 5: end of central contraction; 6: central D-exhaustion, 7: H-ignition, 8:
      ZAMS}
   \label{diag5}
\end{figure}

\subsection{Sensitivity of the birthline to the accretion rate \label{sensibilite}}

Figure~\ref{diag5} shows the relation between the radius and the mass for stars on various
theoretical birthlines.
The noticeable point is that the higher the accretion rate, the more deuterium is available for burning,
the larger the radius and the higher the birthline in the HR diagram.
\begin{displaymath}
   \dot M_{\mathrm{accr}}\,\nearrow\,\Longrightarrow\,\textrm{More D-burning}\,\Longrightarrow \
   \textrm{Bigger R}\,\Longrightarrow\,\textrm{More luminous birthline}
\end{displaymath}

\noindent
This sensitivity of the birthline to $\dot M_{\mathrm{accr}}$ offers an interesting constraint
on the accretion rates. This was discussed with some details by Norberg \& Maeder
(2000). They assume accretion rates of the form
\begin{equation}
   \dot M_{\mathrm{accr}}=A\,(M/M_{\sun})^\varphi
   \label{form4}
\end{equation}
and showed that values $A \approx 10^{-5}\,M_{\sun} \cdot {\mathrm{yr}}^{-1}$
 and $\varphi \approx 1.0 -
1.5$ are leading to the best fit of the birthline on a large set of observational data for
T-Tauri stars and Ae/Be Herbig stars in the HR diagram.

\begin{figure}[!htb]
   \begin{center}
   \includegraphics[width=0.667 \hsize]{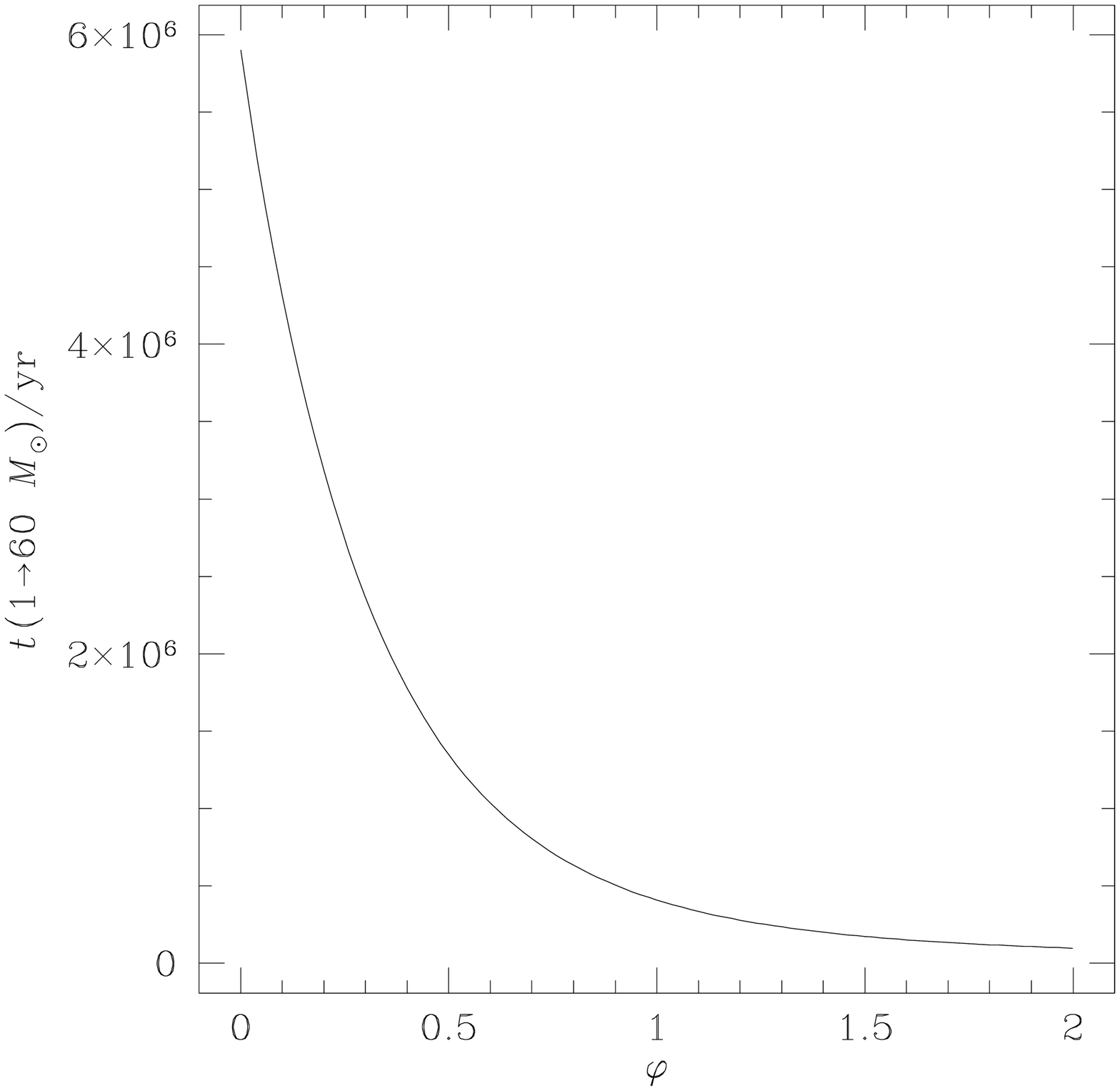}
   \end{center}
   \caption{Time needed by a star to grow from $1$ to $60\, M_{\sun}$ at a rate
       specified by expression~(\ref{form4}), as a function of $\varphi$ with
       $A=10^{-5}\,M_{\sun} \cdot {\mathrm{yr}}^{-1}$}
   \label{diag24}
\end{figure}

The lifetimes $t_{\mathrm{PMS}}$ of the pre--MS phase also depend very much on the accretion
rates. Figure~\ref{diag24} shows the duration $t_{\mathrm{PMS}}$ as a function of the
exponent $\varphi$ in expression~(\ref{form4}).
Note that in Tables~1, 2 \& 3 of Norberg \& Maeder (2000) there is misprint of the decimal
point, the ages should be divided by $10$. The rest of numbers in the tables
are correct.
Figure~\ref{diag24} shows that in order to have pre--MS lifetime of the order of $0.5 \cdot 10^6\,{\mathrm{yr}}$ or less,
it is necessary to have $\varphi \ga 1.0$ If this requirement
is relaxed to $10^6\,{\mathrm{yr}}$, one only needs $\varphi \ga 0.6$
These results support the idea that we need accreting rates growing relatively fast with
the actual stellar mass.

\subsection{Relation with the outflows observed by Churchwell and
Henning}

Massive outflows have been identified from radio and IR observations by Churchwell
(1998) and Henning at al. (2000). The remarkable fact is that there
is a relation between the stellar luminosity and the outflow rate $\dot
M_{\mathrm{out}}$ of the form $\dot M_{\mathrm{out}} \propto L^{0.75}$. According to a
paper presented by Beuther et al. at this meeting, the relation observed by Churchwell and Henning et
al. could
rather be an upper envelope of the $\dot M_{\mathrm{out}}$ distribution as a function of 
luminosity. If we adopt a typical mass luminosity relation for stars on the ZAMS in the range
$2-85\,M_{\sun}$, we get 
\begin{equation}
   \dot M_{\mathrm{out}} \approx
   1.5\,10^{-5}\,(M/M_{\sun})^{1.5}\,M_{\sun} \cdot {\mathrm{yr}}^{-1}.
   \label{form6}
\end{equation}

\noindent
If we assume that a fraction $f$ of the infalling material is accreted by the star
and $(1-f)$ is going into the outflow, we get that a fraction $\tilde f=f/(1-f)$ of the
outflow mass is effectively accreted, {\it i.e.}
\begin{equation}
   \dot M_{\mathrm{accr}}=\tilde f\, \dot M_{\mathrm{out}}.
   \label{form5}
\end{equation}
This implies that the accretion rates grow swiftly with the stellar mass, like $M^{1.5}$.
This result is quite consistent with what we have found in \S~\ref{sensibilite}.
Figure~\ref{diag3} shows numerical models by Behrend \& Maeder (2001)
made with equation (6) and various values of $\tilde f$, 
where $\dot M_{\mathrm{out}}$ is taken from data of Churchwell (1998) and Henning et al. (2000).
We see that models
between those with $\tilde f=0.5$ and $0.176$ beautifully reproduce the upper envelope of
the observations. A value $\tilde f=0.5$ means that one third of the infalling material is
effectively accreted, $\tilde f=0.176$ means $15\%$. These values are those suggested
respectively by the models by Shu et al. (1998) and by the observations by Churchwell
(1998).

\begin{figure}[!htb]
   \plotone{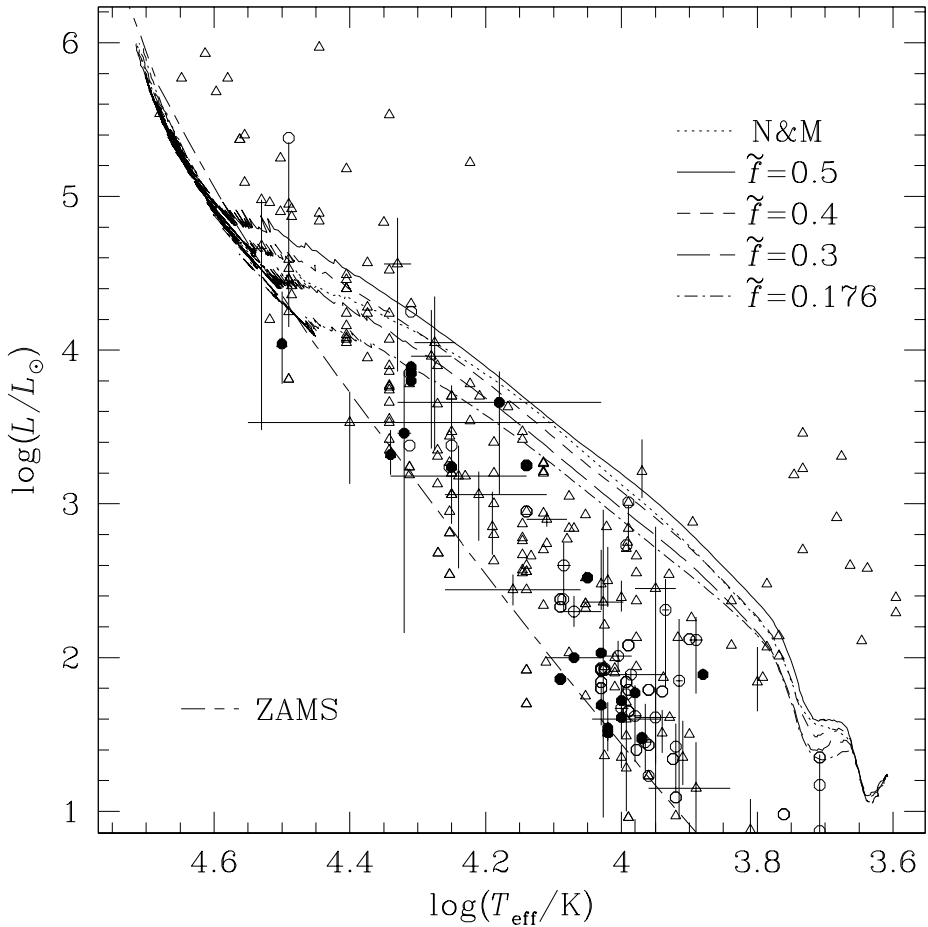}
   \caption{Birthlines in the HR diagram for various values of $\tilde f$
   from Behrend \& Maeder (2001). Values $\tilde f=0.5$ and
      $0.176$ correspond to $1/3$ and $15\%$ of the infalling
 material accreted. A minimum value of $10^{-5}$ 
M$_{\sun} \cdot {\mathrm{yr}}^{-1}$ is taken for low mass stars.
  The observations are those collected by Norberg \& Maeder (2000);
the birthline proposed by these  authors is indicated by N\& M.}
   \label{diag3}
\end{figure}

\begin{table}[!htb]
   \caption[]{Properties of stars on the birthline with $\tilde{f}=0.176$:
   age in units of $10^3\,{\mathrm{yr}}$, luminosity, mass, effective temperature, radius, relative upper radius of the
   convective zone, ratio of disc to star luminosities. From models by Behrend \& Maeder
   (2001).}
   \begin{center}
      $\begin{array}{ccccccc}
       \hline
          t     & \log (L/L_{\sun}) & M/M_{\sun} & T_{\mathrm{eff}}/{\mathrm{K}} & R/R_{\sun} &
          R_{\mathrm{conv}}^{\mathrm{sup}} / R & L_{\mathrm{disc}}/L \\[1mm]
       \hline  
       70.03  & 1.241 & 0.7003 &  4067   &  8.40  & 1.000 & 3.80 \\
       80.94  & 1.099 & 0.8094 &  4183   &  6.74  & 1.000 & 6.41 \\
       90.32  & 1.050 & 0.9032 &  4276   &  6.10  & 1.000 & 8.27 \\
       110.6  & 1.142 &  1.106 &  4402   &  6.39  & 1.000 & 8.41 \\
       128.3  & 1.249 &  1.283 &  4460   &  7.05  & 1.000 & 7.97 \\
       145.4  & 1.331 &  1.454 &  4515   &  7.56  & 1.000 & 7.67 \\
       177.1  & 1.395 &  1.771 &  4615   &  7.79  & 1.000 & 9.04 \\
       197.2  & 1.387 &  1.972 &  4731   &  7.34  & 1.000 & 10.7 \\
       300.3  & 1.334 &  3.003 &  5060   &  6.04  & 1.000 & 20.6 \\
       397.5  & 1.434 &  3.975 &  5354   &  6.05  & 1.000 & 23.8 \\
       455.4  & 2.465 &  5.014 &  7579   &  9.89  & 0.000 & 8.04 \\
       471.0  & 2.959 &  5.985 & 10.41e3 &  9.26  & 0.000 & 6.49 \\
       479.6  & 3.425 &  7.022 & 14.18e3 &  8.54  & 0.000 & 5.24 \\
       484.4  & 3.746 &  8.002 & 18.06e3 &  7.62  & 0.000 & 4.86 \\
       488.0  & 3.992 &  9.038 & 22.36e3 &  6.60  & 0.000 & 4.87 \\
       490.7  & 4.124 &  10.01 & 25.96e3 &  5.69  & 0.014 & 5.49 \\
       493.2  & 4.154 &  11.02 & 27.95e3 &  5.09  & 0.066 & 6.54 \\
       495.6  & 4.176 &  11.99 & 29.05e3 &  4.83  & 0.117 & 7.72 \\
       500.7  & 4.155 &  14.01 & 29.18e3 &  4.67  & 0.145 & 8.77 \\
       506.9  & 4.474 &  17.03 & 33.81e3 &  5.03  & 0.153 & 7.82 \\
       511.6  & 4.608 &  19.98 & 35.49e3 &  5.33  & 0.153 & 6.62 \\
       517.8  & 4.863 &  24.98 & 38.69e3 &  6.01  & 0.173 & 5.41 \\
       522.5  & 5.046 &  29.88 & 40.87e3 &  6.64  & 0.182 & 4.72 \\
       526.8  & 5.219 &  35.20 & 43.11e3 &  7.29  & 0.199 & 4.05 \\
       530.3  & 5.359 &  40.46 & 44.77e3 &  7.94  & 0.209 & 3.61 \\
       533.2  & 5.465 &  45.24 & 46.09e3 &  8.47  & 0.215 & 3.31 \\
       536.1  & 5.567 &  50.56 & 47.17e3 &  9.08  & 0.225 & 3.03 \\
       540.8  & 5.719 &  60.56 & 48.65e3 & 10.17  & 0.237 & 2.60 \\
       544.7  & 5.840 &  69.98 & 49.78e3 & 11.17  & 0.251 & 2.38 \\
       548.7  & 5.952 &  80.65 & 50.56e3 & 12.32  & 0.261 & 2.15 \\
       \hline
      \end{array}$							  
   \end{center}
   \label{tab1}
\end{table}

Table~\ref{tab1} shows some properties of the models by Behrend \& Maeder (2001). 
The disc luminosity is taken as
$L_{\mathrm{disc}}={G\,M\,\dot M_{\mathrm{inflow}} \over 2\,R}$ with $\dot
M_{\mathrm{inflow}}=\dot M_{\mathrm{out}}+\dot M_{\mathrm{accr}}$ ({\it cf}. Hartmann
1998). Of course there is a proportionality  factor which is uncertain and thus 
these values for $L_{\mathrm{disc}}$ are only indicative. 
Table~\ref{tab1} shows that the last stages of massive
star accretion go very fast. For example, in Table~\ref{tab1} the time for the evolution from
$15$ to $80\,M_{\sun}$ is $4.7\ 10^4\,{\mathrm{yr}}$. For $\tilde f=0.5$, this time would be
only $1.5\,10^4\,{\mathrm{yr}}$. Of course, for different values of $\varphi$ as in
equation~(\ref{form4}), we would have different timescales as suggested by Figure~\ref{diag24}.
As we do not know
exactly what is the value of $\varphi$ and $\tilde f$, this means that the lifetime for
the fast phases of massive star accretion, {\it i.e.}~from 
$15\,M_{\odot}$ to $80\,M_{\odot}$ are still rather uncertain, somewhere between a
few $10^4\,{\mathrm{yr}}$ and a few $10^5\,{\mathrm{yr}}$.

\section{Models and discussion}

\subsection{Physics feasibility of the high accretion rate \label{faisable}}

The accretion rates given by equations~(\ref{form4}) and~(\ref{form5}) are remarkably well
located in the stable zone derived by Wolfire \& Cassinelli (1987). For rates lower than a
certain limit, the momentum in the radiation pressure is sufficient to prevent the
accretion. For rates higher than some limit, the luminosity created by the shock is
supra-Eddington. It has been shown by Nakano (1989) that for a non-spherical collapse the
stability conditions  are much less severe than those given  by Wolfire \& Cassinelli.
In particular, Nakano (1989) suggests that, even for the most massive stars,
radiation cannot stop the accretion process and as a consequence he suggests 
that the maximum stellar mass is determined by fragmentation rather by the
accretion process. Certainly, both fragmentation and accretion
are essential in shaping the IMF and determining the maximum mass.
Indeed, if most of the infalling material is diverted in the outflows,
this implies that only a tiny fraction of the  initial fragmentats finally contributes to the IMF.

We also point out  that the various accretion models do not account for the
anisotropy of the radiation field of massive rotating stars, which would 
favour anisotropic accretion in pre--MS stages and anisotropic 
mass loss in post--MS phases ({\it cf}. Maeder \& Desjacques 2001).
Indeed, the lower $T_{\mathrm{eff}}$ in the equatorial regions 
of a rotating massive pre--MS object should 
very much favour the accretion  on these objects. Joined to the
above results of Nakano (1989), this supports the idea
that accretion at heavy rates of the order of
$\dot M_{\mathrm{accr}} \approx 10^{-3}\,M_{\sun} \cdot {\mathrm{yr}}^{-1}$
is quite possible theoretically.

\subsection{Dependence of the accretion rate on the central mass \label{depend}}

The location of the birthline, the timescales and the observations of the outflows support
accretion rates $\dot M_{\mathrm{accr}} \approx 10^{-5}\,(M/M_{\sun})^\varphi$ with
$\varphi \ga 1$ as shown above. Globally, we may understand a relation of this kind since
for larger masses the central potential is much larger and attracts more matter. The
accretion must go like
\begin{displaymath}
   \dot M_{\mathrm{accr}} \propto {M \over t_{\mathrm{ff}}} \propto M\,(G\,\bar
   \rho)^{1/2} \propto (M/R)^{3/2}
\end{displaymath}
where $M$ is the mass of the initial cloud of average density $\bar \rho$,
$t_{\mathrm{ff}}$ is the free fall time. The radius of the initial subcondensation may
vary with their mass in different ways according to different authors. 
Larson (1981) has
suggested for small condensations (Larson's scaling)
\begin{displaymath}
   \bar \rho \propto {1 \over R}.
\end{displaymath}
If so, $R \propto M^{1/2}$ and thus $\dot M_{\mathrm{accr}} \propto M^{3/4}$. In this
meeting, Johnstone (2001) has suggested that the density of the subcondensations is
constant (Johnstone's scaling)
\begin{displaymath}
   \bar \rho \propto \textrm{const}.
\end{displaymath} 
In this case, $R \propto M^{1/3}$ and we have
\begin{equation}
   \dot M_{\mathrm{accr}} \propto M.
\end{equation}
Interestingly enough,
Johnstone's scaling implies accretion rate growing almost linearly with the mass as
suggested by the observations quoted above. The above arguments are rather schematic and
ignore many effects like rotation, accretion disc, outflows, etc. Nevertheless, the general
trend they show must be roughly correct.
Very detailed numerical models are now in progress in Behrend's thesis. With finite
elements grids, which allow us to follow more closely the interesting parts of the models, the
accretion with rotation is modelized, the inflow of the matter through the disc followed
as well as the central accretion and outflows. These models predict the accretion rates
and should be combined with interior models  of the central body.

\subsection{Formation of binaries}

An apparent difficulty of the accretion scenarios (it is sometimes presented as objection) is
the question on how massive binary stars do form. The question is especially important
because the fraction of binaries and multiple stars is very high as shown by Zinnecker
(this meeting). Here, we would like to suggest some possibilities of preferential binary
formation among the OB-stars.

Let us consider a molecular cloud, which is collapsing to give birth to a young star
cluster. In the process of cloud fragmentation, some cores may be bound gravitationally, some do not. Each
core, bound or not, will accrete matter from its own appropriate Jeans radius.
However, the bound or multiple cores will also, in addition to their current
accretion, experience a Bondi-Hoyle accretion particularly if
their Bondi-Hoyle radius $b$
\begin{equation}
   b=2\,{G\,M \over v^2}
   \label{form8}
\end{equation}
is greater than the semi-major axis of the supposed double system. In this case, the velocity $v$
appearing in equation~(\ref{form8}) is some combination of the orbital velocity in the
double system and of the velocity of the double system in the cluster.
Thus, double and multiple systems have a higher potential reservoir of matter for
accretion and their accretion rates $\dot M_{\mathrm{accr}}$ should also be larger than that for
single stars.

This means that double and multiple cores, by their larger reservoir and higher
$\dot M_{\mathrm{accr}}$ will have more chance to move to the top of the mass scale (the IMF)
and they will do it faster than the other single objects. This may thus lead to a higher
number of double and multiple systems among OB-stars in young clusters.
Since mass is accreted on the protobinary system, the semi-major axis will decrease
and the orbital periods will also decrease. It is to be examined whether an orbital separation corresponding
to the initial tidal radius within the cluster finally leads to periods
$P<10\,{\mathrm{d}}$, as so frequently observed in massive O-stars.

In conclusion, we see that many observations and  theoretical considerations
effectively support the accretion scenario for massive stars. The high frequency of binaries
among O--type stars might even be an additional argument in favour of the accretion scenario.


\end{document}